%% file: 25ESSERC_LDPC.tex
\documentclass[conference,10pt,twoside,twocolumn]{IEEEtran}

\usepackage{cite}
\usepackage[T1]{fontenc}
\usepackage{graphicx}
\usepackage{amssymb}
\usepackage{amsmath}
\usepackage{amsthm}
\usepackage{subcaption}
\usepackage{caption}
\usepackage{booktabs} 
\usepackage{multirow}
\usepackage{microtype}
\usepackage{balance}
\usepackage{xcolor}
\usepackage{cite}
\usepackage{microtype}
\usepackage{threeparttable}

\input{macros/vmr-symbols-vecbold}
\input{macros/standard-macros}

\input{macros/defs}
\safemath{\Tran}{\textnormal{T}}
\safemath{\Herm}{\textnormal{H}}

\IEEEoverridecommandlockouts
\allowdisplaybreaks 

\begin{document}
\bstctlcite{IEEEexample:BSTcontrol}
\title{A 14\,ns-Latency 9\,Gb/s 0.44\,$\textnormal{mm}^\textnormal{2}$ 62\,pJ/b Short-Blocklength LDPC Decoder ASIC in 22FDX}

\author{\IEEEauthorblockN{Darja Nonaca$^*$, J\'er\'emy Guichemerre$^*$, Reinhard Wiesmayr$^*$, Nihat Engin Tunali$^\diamondsuit$, and Christoph Studer$^*$}\\[-0.20cm]
\IEEEauthorblockA{\em $^*$Department of Information Technology and Electrical Engineering, ETH Zurich, Switzerland; $^\diamondsuit$Independent Researcher}\\[-0.65cm]

\thanks{This work has received funding from the Swiss State Secretariat for Education, Research, and Innovation (SERI) under the SwissChips initiative. The authors thank GlobalFoundries for providing silicon fabrication through the 22FDX University Program. 
The authors also thank Oscar Casta\~neda for his assistance with the ASIC design flow. Contact author: D. Nonaca (e-mail: dnonaca@iis.ee.ethz.ch).}
}

\maketitle  

\begin{abstract}
Ultra-reliable low latency communication (URLLC) is a key part of 5G wireless systems. Achieving low latency necessitates codes with short blocklengths for which polar codes with successive cancellation list (SCL) decoding typically outperform message-passing (MP)-based decoding of low-density parity-check (LDPC) codes. However, SCL decoders are known to exhibit high latency and poor area efficiency. 
In this paper, we propose a new short-blocklength multi-rate binary LDPC code that outperforms the 5G-LDPC code for the same blocklength and is suitable for URLLC applications using fully parallel MP.
To demonstrate our code's efficacy, we present a 0.44\,mm\boldmath{$^2$} GlobalFoundries 22FDX LDPC decoder ASIC which supports three rates and achieves the lowest-in-class decoding latency of 14\,ns while reaching an information throughput of 9\,Gb/s at 62\,pJ/b energy efficiency for a rate-1/2 code with 128-bit blocklength. 
\end{abstract}

\input{chapters/intro}

\section{LDPC Decoding} 
\label{sec:ldpc}

We consider forward error correction with binary block codes, where  $\vecb\in\{0,1\}^k$ is the information bit vector of length $k$ to be transmitted and $\vecc= \matG \vecb\in\{0,1\}^n$ the resulting codeword of length $n$. Here, $\matG$ is the generator matrix, typically derived from the parity-check (PC) matrix $\matH$ for which every codeword satisfies $\bH\bmc = \bZero$.
Instead of transmitting $\bmc$, one often transmits $\vecc'\in\{0,1\}^{n'}$ with $n'\leq n$ by \textit{puncturing} (removing) a fixed set of bits from $\vecc$ that will later be retrieved by the decoder.
With puncturing, the code rate is given by $R=k/n'$.

The coded and punctured bits $\vecc'$ are mapped to the symbol sequence $\vecx$ and transmitted over a channel. The receiver then obtains ${\vecy=h(\vecx)}$, where $h$ models the  channel.
At the receiver, the entries $y_i$ of $\vecy$ are used to compute the a-posteriori likelihood of each bit $c_i$, from the codeword $\vecc$, described by the log-likelihood ratio (LLR) $\ell_i = \log\!\Big(\frac{\Prob(c_i = -1|y_i)}{\Prob(c_i = +1|y_i)}\Big)$; the LLR values associated with the punctured bits are set to zero. 

In what follows, we focus on LDPC codes~\cite{RG62,MK99}, which can be represented as a bipartite graph consisting of variable nodes (VNs) associated with the LLR values, and check nodes (CNs) associated with the parity checks (corresponding to the rows of~$\bH$).
LDPC codes can be decoded using MP~\cite{MK99} by iteratively propagating beliefs along the edges of the bipartite graph in an iterative fashion. 
We implement the flooding MP schedule, which alternates between updating the messages associated with all CNs and all VNs, together with the edge-adaptive normalized min-sum approximation (NMS)~\cite{Chen02} 
\begin{align}
\footnotesize
   \ell\Bigl(c_j=\mkern-16mu\mathop{\bigoplus}_{k\in N(i)\backslash{j}}\mkern-16mu c_k\Bigr)\approx \alpha_{ij}\times\mkern-12mu\prod_{k\in N(i)\backslash{j}}\mkern-12mu\sign{\ell_{k}}\;\times\min_{k\in N(i)\backslash{j}}\abs{\ell_{k}}. \label{eq:check_node_eq}
\end{align}
Here, $N(i)$ is the set of the VNs connected to the $i$\,th CN and $\alpha_{ij}$ is a predefined scaling factor associated with the edge connecting the $i$\,th CN and the VN associated to~$c_j$. 
The scaling factors are obtained off-line through machine learning training as introduced in~\cite{EN18}, with additional fine-tuning utilizing SNR deweighting and a BLER loss function as proposed in~\cite{Wiesmayr2023}.\footnote{As in~\cite{EN18}, we optimize one set of trainable CN-to-VN edge-weights for the scaling parameters $\alpha_{ij}$ that are shared among all MP iterations.}
We terminate the decoding process either if a maximum number of iterations $I_\textnormal{max}$ is reached or if a valid codeword is found.

\section{Custom Short-Blocklength QC-LDPC Code}
To achieve low error rates with MP-based decoding, we propose a custom short-blocklength binary quasi-cyclic (QC)~\cite{MF04} LDPC code derived from the family of rate-compatible protograph-based AR4A codes~\cite{DD09}. The parity-check matrix of the rate $3/4$ code is represented by the~protograph 
\begin{align}
\footnotesize
\mathbf{H}_\textnormal{proto} = \left[\begin{array}{ccccccccc}
0 & 0 & 0 & 0 & 2 & 0 & 2 & 0 & 0 \\
3 & 1 & 3 & 1 & 0 & 1 & 3 & 2 & 0 \\
1 & 3 & 1 & 3 & 0 & 3 & 1 & 0 & 2
\end{array}\right]\!,
\end{align}
where each entry represents the constraints on connectivity of the variable nodes and check nodes.
For example, the variable nodes in the first column would only be connected to three check nodes in the middle row and a single check node in the last row.
Code construction consists of two stages:
In the first stage, each entry in the protograph matrix~$\bH_\textnormal{proto}$ is expanded (lifted) by a lifting factor $Z=4$ with the progressive edge growth (PEG) algorithm~\cite{XYH01}.
At this stage, the matrix is not QC, but has increased girth for improved MP decoding performance. 
In the second stage, the PEG expanded PC matrix is lifted by $Z=8$ with the approximate cycle extrinsic (ACE) message degree algorithm~\cite{TT03}, by setting $d_\textnormal{ACE}=3$ and $\eta_\textnormal{ACE}=13$; this ensures that length $2\times d_\textnormal{ACE}$ cycles have an ACE message degree of $13$.
The larger the ACE message degree, the better a cycle is connected to the rest of the graph.
The resulting PC matrix consists of cyclically shifted identity matrices of size $Z=8$ (cf.~\fref{fig:code_matrix}). Combining the PEG and ACE algorithms is known to achieve near Shannon performance~\cite{DD09}.

\begin{figure*}[t]
   \centering
    \begin{subfigure}{0.32\textwidth}
        \centering
        \includegraphics[width=\textwidth]{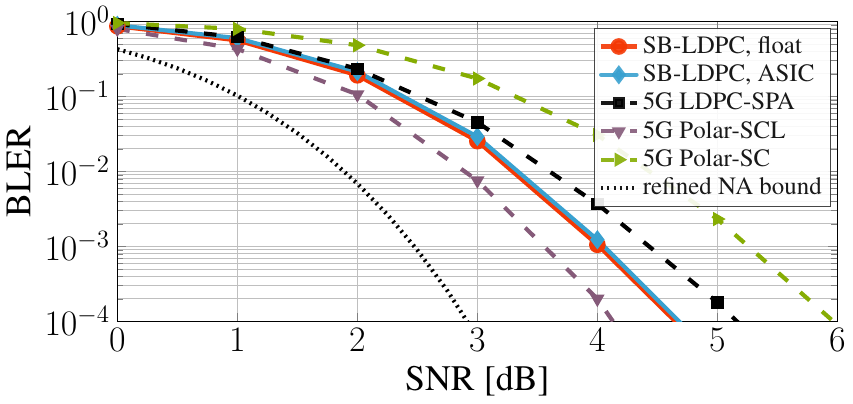}
        \caption{$R=1/2, n = 160, n' = 128$}
        \label{fig:subfig1}
    \end{subfigure}
    \hfill
    \begin{subfigure}{0.32\textwidth}
        \centering
        \includegraphics[width=\textwidth]{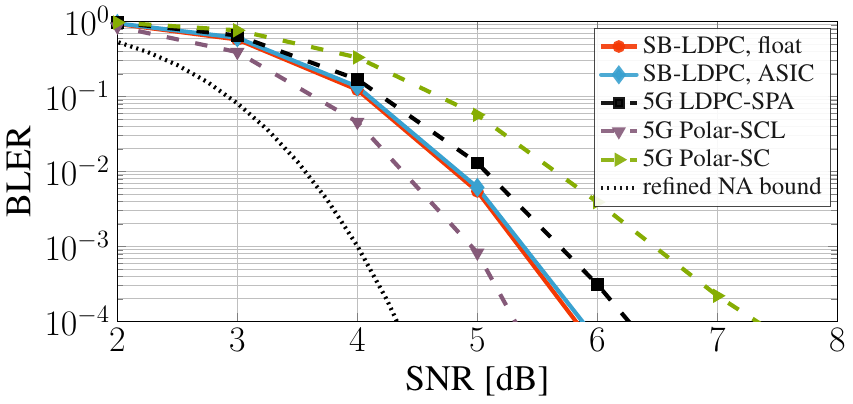}
        \caption{$R=2/3, n = 224, n' = 192$}
        \label{fig:subfig2}
    \end{subfigure}
    \hfill
    \begin{subfigure}{0.32\textwidth}
        \centering
        \includegraphics[width=\textwidth]{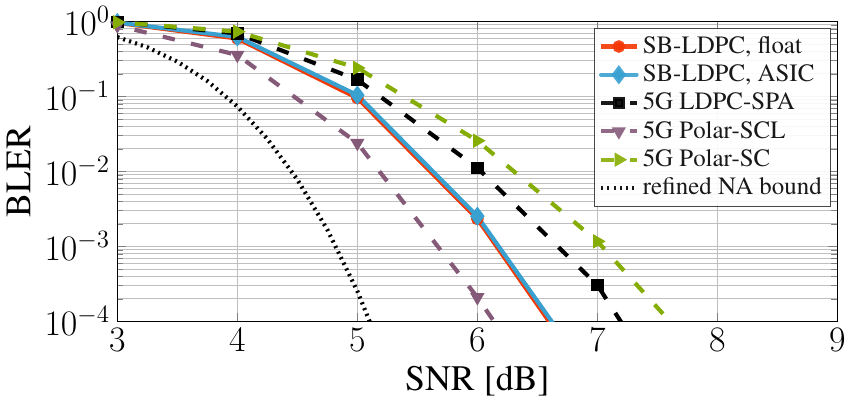}
        \caption{$R=3/4, n = 288, n' = 256$}
        \label{fig:subfig3}
    \end{subfigure}
    \caption{BLER comparison at a fixed blocklength and rate of the proposed short blocklength (SB) LDPC code using the update rule in \fref{eq:check_node_eq} with the 5G LDPC code using the SPA update rule with the same number of decoding iterations $I_\textnormal{max}=10$, and with 5G polar codes with SC and SCL decoding using list size 8. Our SB-LDPC code is $1.5$\,dB away from the refined NA bound and outperforms the 5G LDPC code by $0.3-0.5$\,dB at $0.1\,\%$ \;BLER for all three rates.}
    \label{fig:bler}
    \vspace{-0.15cm}
\end{figure*}

Our proposed code is depicted in~\fref{fig:code_matrix} and supports three different code lengths for three different rates: $R=1/2$, $R=2/3$, and $R=3/4$. To adjust the code rate (and length), the columns of the PC matrix $\bH$ are progressively removed. For $R=3/4$, the entire PC matrix $\bH$ is used; for $R=2/3$ and $R=1/2$, the first $64$ and $128$ columns are removed, respectively.  
In this code, we puncture the information bits $[1\!:\!32]$, $[65\!:\!96]$, and $[129\!:\!160]$ for rates $R=1/2$, $R=2/3$, and $R=3/4$, respectively.
This procedure leads to the code lengths $(k,n')=(64,128)$, $(k,n')=(128,192)$, and $(k,n')=(192,256)$, for rates $R=1/2$, $R=2/3$, and $R=3/4$, respectively.
To demonstrate the efficacy of our short-blocklength LDPC (SB-LDPC) code, we simulate a memoryless, real-valued, AWGN channel with binary phase-shift keying. 
\fref{fig:bler} evaluates the block error rate (BLER) of the proposed SB-LDPC code with ANMS decoding for rates $1/2$, $2/3$, and $3/4$.
We compare the BLER of our decoding method applied to our SB-LDPC code with the theoretical bound called refined normal approximation (NA)~\cite{durisi20-11a}, the 5G LDPC code decoded with the sum-product MP algorithm (SPA), and the 5G polar codes with SC and SCL decoding.
For every LDPC decoder, we use the flooding scheduling with $I_\textnormal{max}=10$ iterations; for polar SCL, we use a list size $8$.
\fref{fig:bler} reveals that for all three rates, our code and decoding scheme exhibit a loss of approximately $1.5$\,dB relative to the refined NA bound and $0.5$\,dB from 5G polar codes with SCL decoding. 
Depending on the rate, we achieve a $0.3$ to $0.5$\,dB gain over 5G LDPC codes with SPA decoding and a $0.75$ to $1.25$\,dB gain over 5G polar codes with SC decoding.%
\fref{fig:bler} also demonstrates that our fixed-point hardware design exhibits virtually no BLER loss.
%
\input{chapters/vlsi}
\begin{figure}[tp]
\centering
\includegraphics[width=0.9\columnwidth]{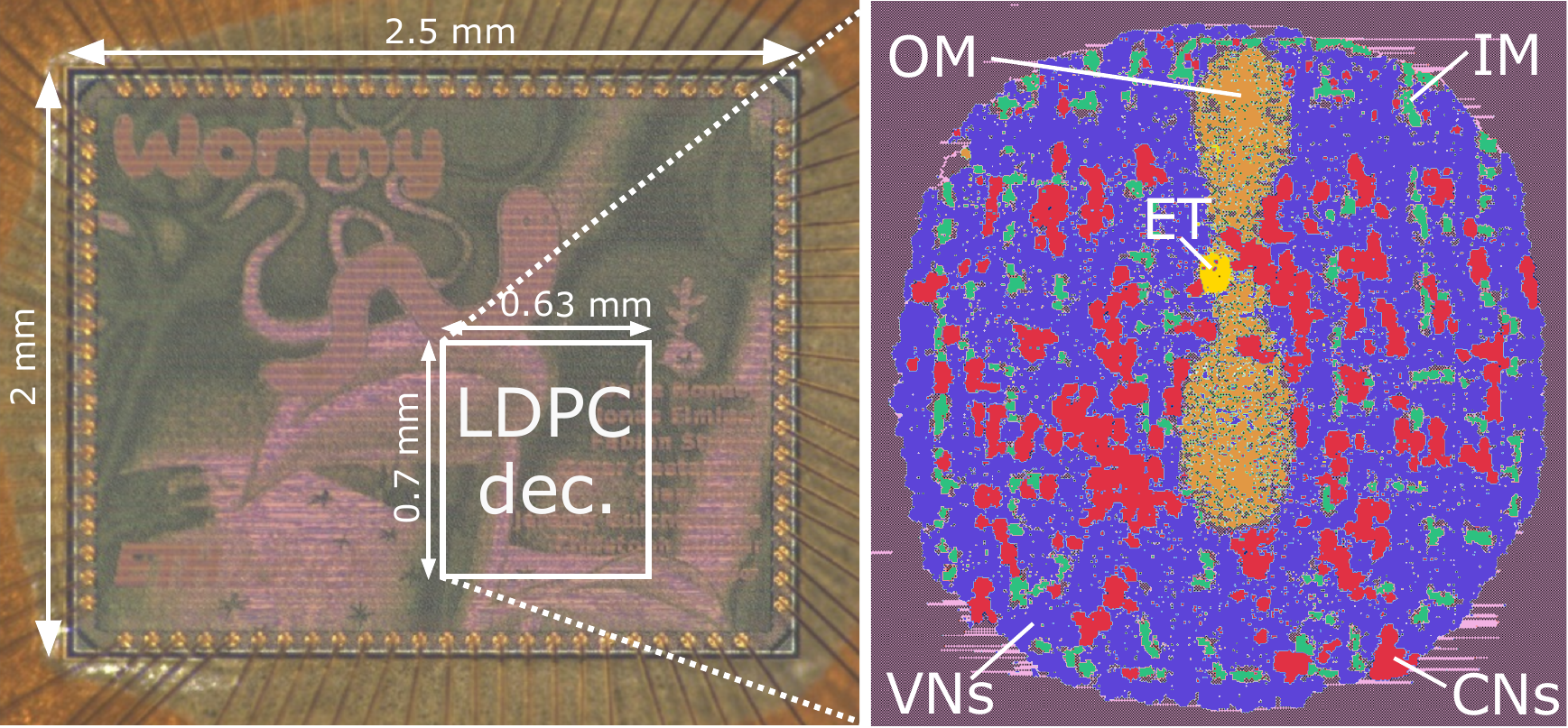}
\caption{ASIC micrograph (left) containing our SB-LDPC decoder and detailed area breakdown (right). OM, IM, and ET are output LLR memory, input LLR memory, and early termination block, respectively. Each VN block contains PUs, one for each VN-CN connection from the given VN block.}
\label{fig:chip}
\vspace{-0.1cm}
\end{figure}

\section{Implementation Results and Comparison}\label{sec:results}

\begin{table*}[t]  
\setlength{\tabcolsep}{4pt} 
    \caption{ASIC measurement results and comparison with other decoder designs.}
    \label{tbl:comp}
   \vspace{-0.1cm}
    \centering
    \resizebox{1.00\textwidth}{!}{ 
    \begin{tabular}{@{}lccc|ccccccc@{}}
        \toprule
        ~ & \multicolumn{1}{c}{This work} & {This work} &{This work}  & Ghanaatian & Zhang & Milicevic  & Verma & Teng & Giard & Kam\\
        ~ & \multicolumn{1}{c}{$R=1/2$}& {$R=2/3$} & {$R=3/4$} & \cite{RG18} & \cite{ZZ10}& \cite{MM18} & \cite{AV24} & \cite{CT21} &\cite{PG17} &\cite{DK24}\\
        \hline
        Code/Algorithm & & LDPC-ANMS & &LDPC-LUT & LDPC-OMS &LDPC-MS & LDPC-OMS & Polar RNN-BP & Polar-SCL & BOSS\\
        \hline
        $(k,n')$ & (64,128) & (128,192) & (192,256)&(1723,2048)&(1723,2048)&(336,672)&(352,528) & (128,256) & (512,1024) & (15,128)\\
         \hline
        Rate~$R$      & 0.5         & 0.67         & 0.75       & 0.84             & 0.84 & 0.5 & 0.67 & 0.5 & 0.5 & 0.12\\
        \hline
        Max.~iterations~$I_\textnormal{max}$ &10 & 10 & 10&5&14&10&10&5&- &-\\
        \hline
        Min.~SNR\;[dB]\;@BLER$=10^{-3}$ & 4 & 5.6 & 6.2 & 6.90$^\textit{a}$ & 6.41$^\textit{a}$ & 4.38$^\textit{a}$ &2.33$^\textit{i}$ & BER reported & 2.4$^\textit{a}$ & -2.55$^\textit{a}$  \\
       \hline
        Technology ($L$)\;[nm] & 22& 22 & 22 & 28 & 65 & 28 &110 & 40 & 28 & 28\\
       \hline
        Fabricated? &yes &yes& yes &no &yes &yes &yes & yes & yes & yes\\
       \hline
        Supply voltage ($V_s$)\;[V] &0.8 & 0.8 & 0.8  &1&1.2&0.9&1.2 & 0.9 & 0.9 & 0.95\\
        \hline
        Area\;[$\text{mm}^2$] &0.44 & 0.44 &0.44&16.2& 5.05&1.99&1.96 & 0.18 &0.44 & 0.37\\
        \hline
        $f_\textnormal{max}$ no ET / ET\;[MHz] & 1452 / 1356& 1246 / 809 & 1142 / 776 & 862 &700&202&72.7& 225 & 308 & 590\\
        \hline
        Power\;@$f_\textnormal{max}$ no ET / ET [mW]
       & 575 / 296 & 636 / 218 &699 / 236&  13350 & - / 2800& 408 / 283 &150 & 12.8 & 23.3 & 33.3\\
        \hline
        Inf.~throughput\;@$f_\textnormal{max}$ and $I_\textnormal{max}$\;[Gb/s]& 9.29$^\textit{b}$&15.94$^\textit{b}$&21.92$^\textit{b}$ & 494.68 & 40.13$^\textit{d}$& 3.39 & 1.11 & 0.41 & 0.06 & 0.68\\
        \hline
        Latency\;@$I_\textnormal{max}$\;[ns] &13.78$^\textit{b}$&16.06$^\textit{b}$&17.52$^\textit{b}$&69.6 & 137 &793 &120 & 310 & 7820 & 21.9\\
        \hline
        Area efficiency\;[Gb/s/$\text{mm}^2$]&21.12$^\textit{b}$&36.24$^\textit{b}$&49.83$^\textit{b}$ &30.53&7.94$^\textit{d}$&1.70& 0.56 & 2.28 &0.14 & 1.84\\
        \hline
        Energy efficiency\;@$I_\textnormal{max}$\;[pJ/b] & 61.88$^\textit{b}$ & 39.88$^\textit{b}$ & 31.88$^\textit{b}$ & 26.99 & 69.78$^\textit{c,d}$ & 120.36$^\textit{b}$ & 134.74 & 31.11 & 355.73 & 48.66\\
        \hline
        \hline
        Scaled area efficiency$^\textit{e}$\;[Gb/s/$\text{mm}^2$]&21.12$^\textit{b}$ & 36.24$^\textit{b}$ & 49.83$^\textit{b}$& 62.95 & 204.95$^\textit{d}$ &3.51& 71 & 13.74 & 0.30 & 3.81\\
        \hline
        Scaled energy efficiency$^\textit{f}$\;[pJ/b]&61.88$^\textit{b}$&39.88$^\textit{b}$&31.88$^\textit{b}$&13.57&10.50$^\textit{c,d}$& 74.72$^\textit{b}$ & 11.98 & 13.52 & 220.84 & 27.12\\
        \hline
        Scaled inf.~throughput$^\textit{g}$\;@$f_\textnormal{max}$ and $I_\textnormal{max}$\;[Gb/s]& 9.29$^\textit{b}$&15.94$^\textit{b}$&21.92$^\textit{b}$& 629.60& 118.56$^\textit{d}$ & 4.31  & 5.56 & 0.74 & 0.08  & 0.87\\
        \hline
        Scaled latency$^\textit{h}$\;[ns]&\textbf{13.78}$^\textit{b}$ &\textbf{16.06}$^\textit{b}$ &\textbf{17.52}$^\textit{b}$ &54.69&46.37$^\textit{d}$&623.08 & 24 & 170.5 & 6144.29 & 17.21\\
        \bottomrule
    \end{tabular}
    } \\[0.1cm]    \raggedright\hspace{-0.1cm}\footnotesize{    
    \tiny{$^{\textit{a}}$}Conversion from $\textnormal{\textit{E}}_\textnormal{{\fontsize{4pt}{4pt}\selectfont b}}/\textnormal{\textit{N}}_\textnormal{{\fontsize{4pt}{4pt}\selectfont 0}}$ to SNR. %
    \tiny{$^{\textit{b}}$}Considering decoding without ET.
    \tiny{$^{\textit{c}}$}Considering decoding with ET.
    \tiny{$^{\textit{d}}$}Average throughput reported; the average throughout value taken at 5.5\,dB with only 1.5 iterations.
    \tiny{$^{\textit{e}}$}Technology scaling by $\textnormal{\textit{S}}^{\textnormal{{\fontsize{4pt}{4pt}\selectfont 3}}}$, \tiny{$^{\textit{f}}$}  \textit{S}$^{\textnormal{{\fontsize{4pt}{4pt}\selectfont -1}}}$$\textnormal{\textit{U}}^{\textnormal{{\fontsize{4pt}{4pt}\selectfont -2}}}$, \tiny{$^{\textit{g}}$}$\textnormal{\textit{S}}$, and \tiny{$^{\textit{h}}$$\textnormal{\textit{S}}^{\textnormal{{\fontsize{4pt}{4pt}\selectfont -1}}}$}, where $\textnormal{\textit{S = L}}/$22 is the relative dimension to 22\,nm and \textit{U} = $\textnormal{\textit{V}}_{\textnormal{s}}/$0.8 the relative core voltage to 0.8\;V. \tiny{$^{\textit{i}}$}Reported BLER is below the refined NA bound~\cite{durisi20-11a} for the corresponding blocklength and rate.
    }
    
   \vspace{-0.2cm}
\end{table*}

\begin{figure}[tp]
\centering
\hfill
    \begin{subfigure}{0.45\columnwidth}
        \centering
        \includegraphics[width=0.99\columnwidth]{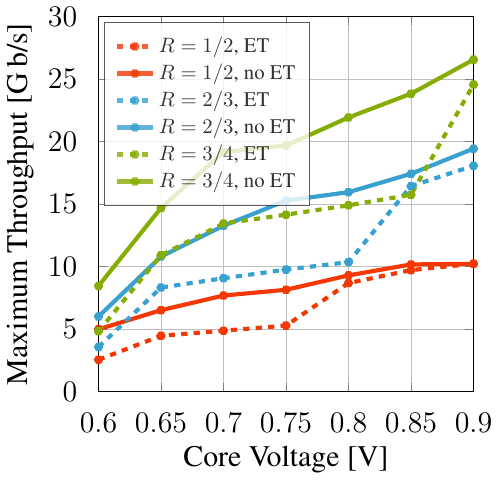}
        \label{fig:max_throughput}
    \end{subfigure}
\hfill
    \begin{subfigure}{0.45\columnwidth}
        \centering
        \includegraphics[width=0.99\columnwidth]{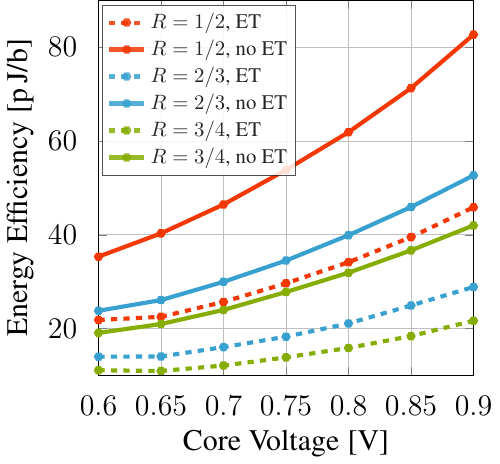}
        \label{fig:energy_eff}
    \end{subfigure}
\vspace{-0.5cm}
\caption{Measured maximum throughput (left) and energy efficiency (right) achieved by our LDPC decoder for the three supported rates. The measurements for each rate were taken at the minimum SNR that achieves $0.1\%$ BLER, i.e., $4$\,dB, $5.4$\,dB, and $6.2$\,dB for rates $R=1/2$,  $R=2/3$, $R=3/4$, respectively.  }
\label{fig:chip_measurments}
\end{figure}

\fref{fig:chip} depicts our multi-project chip fabricated in GlobalFoundries’ 22\,FDX\textsuperscript{\sf\tiny TM} FD-SOI technology. 
The proposed SB-LDPC decoder occupies an area of $0.44$\,$\textnormal{mm}^2$, including the input memory (IM) to store the two input LLR vectors and output memory (OM) for the two output LLR vectors, both of size $7\times288$ bits. 
At nominal $0.8$\,V core supply voltage and ${25}^\circ$\,C, the decoder achieves a maximum clock frequency of $1.452$\,GHz, $1.246$\,GHz, and $1.142$\,GHz for the rates $R=1/2$, $R=2/3$, $R=3/4$, respectively.
Our ASIC achieves higher data rates with longer blocklengths, which activates more VN blocks; the power and critical path leads to a reduced clock frequency $f_\textnormal{max}$.
With $I_\textnormal{max} = 10$, we achieve an information throughput $\theta_k = kf_\textnormal{max}/I_\textnormal{max}$ of $9.29$, $15.94$, and $21.92$\,Gb/s at rates $R=1/2$, $R=2/3$, $R=3/4$, respectively, as shown on the left side of~\fref{fig:chip_measurments}.
ET is in the critical path and, when activated, reduces the throughput. Nonetheless, ET substantially improves energy efficiency; see~\fref{fig:chip_measurments}.

\fref{tbl:comp} compares our SB-LDPC decoder with the state of the art. 
With respect to LDPC decoders for long blocklengths~\cite{RG18,ZZ10}, our ASIC achieves the shortest latency.
With respect to short blocklength codes~\cite{MM18,AV24,CT21,PG17,DK24}, our ASIC achieves the highest information throughput and is the most area efficient except for~\cite{AV24}. 
With respect to the LDPC decoders~\cite{RG18,ZZ10,MM18,AV24} for long and short blocklengths, our custom short-blocklength LDPC code and ASIC achieve lower BLER (considering equal code rates), except for~\cite{AV24}, which reports a BLER below the refined NA bound. 
In summary, our SB-LDPC decoder achieves the shortest latency while being competitive in (scaled) throughput and area efficiency as well as BLER performance. 

\section{Conclusions}

We have proposed a new short blocklength LDPC code along with a decoder ASIC that relies on a parallel message-passing architecture. 
The proposed rate-configurable decoder outperforms existing LDPC decoders for short blocklengths in terms of BLER and achieves a throughput of $9$, $16$, and $22$\,Gb/s for rates $R=1/2$, $R=2/3$, and $R=3/4$, respectively. 
Furthermore, our decoder achieves decoding latencies in the range of $14$\,ns to $18$\,ns---the shortest reported in the literature.

\linespread{0.9475}

\bibliographystyle{IEEEtran}
\bibliography{bib/VIPabbrv,bib/confs-jrnls,bib/publishers,bib/VIP_190331, bib/25ESSERC}

\end{document}

%% file: macros/vmr-symbols-vecbold.tex
\usepackage{amssymb}
\usepackage{amsfonts}
\usepackage{mathrsfs}
\usepackage{xspace}
\usepackage{bm}
\usepackage{upgreek}

\newcommand{\safemath}[2]{\newcommand{#1}{\ensuremath{#2}\xspace}}

\safemath{\bma}{\mathbf{a}}
\safemath{\bmb}{\mathbf{b}}
\safemath{\bmc}{\mathbf{c}}
\safemath{\bmd}{\mathbf{d}}
\safemath{\bme}{\mathbf{e}}
\safemath{\bmf}{\mathbf{f}}
\safemath{\bmg}{\mathbf{g}}
\safemath{\bmh}{\mathbf{h}}
\safemath{\bmi}{\mathbf{i}}
\safemath{\bmj}{\mathbf{j}}
\safemath{\bmk}{\mathbf{k}}
\safemath{\bml}{\mathbf{l}}
\safemath{\bmm}{\mathbf{m}}
\safemath{\bmn}{\mathbf{n}}
\safemath{\bmo}{\mathbf{o}}
\safemath{\bmp}{\mathbf{p}}
\safemath{\bmq}{\mathbf{q}}
\safemath{\bmr}{\mathbf{r}}
\safemath{\bms}{\mathbf{s}}
\safemath{\bmt}{\mathbf{t}}
\safemath{\bmu}{\mathbf{u}}
\safemath{\bmv}{\mathbf{v}}
\safemath{\bmw}{\mathbf{w}}
\safemath{\bmx}{\mathbf{x}}
\safemath{\bmy}{\mathbf{y}}
\safemath{\bmz}{\mathbf{z}}
\safemath{\bmzero}{\mathbf{0}}
\safemath{\bmone}{\mathbf{1}}

\bmdefine{\biad}{a}
\bmdefine{\bibd}{b}
\bmdefine{\bicd}{c}
\bmdefine{\bidd}{d}
\bmdefine{\bied}{e}
\bmdefine{\bifd}{f}
\bmdefine{\bigd}{g}
\bmdefine{\bihd}{h}
\bmdefine{\biid}{i}
\bmdefine{\bijd}{j}
\bmdefine{\bikd}{k}
\bmdefine{\bild}{l}
\bmdefine{\bimd}{m}
\bmdefine{\bind}{n}
\bmdefine{\biod}{o}
\bmdefine{\bipd}{p}
\bmdefine{\biqd}{q}
\bmdefine{\bird}{r}
\bmdefine{\bisd}{s}
\bmdefine{\bitd}{t}
\bmdefine{\biud}{u}
\bmdefine{\bivd}{v}
\bmdefine{\biwd}{w}
\bmdefine{\bixd}{x}
\bmdefine{\biyd}{y}
\bmdefine{\bizd}{z}

\bmdefine{\bixid}{\xi}
\bmdefine{\bilambdad}{\lambda}
\bmdefine{\bimud}{\mu}
\bmdefine{\bithetad}{\theta}
\bmdefine{\biphid}{\phi}
\bmdefine{\bideltad}{\delta}

\safemath{\bmia}{\biad}
\safemath{\bmib}{\bibd}
\safemath{\bmic}{\bicd}
\safemath{\bmid}{\bidd}
\safemath{\bmie}{\bied}
\safemath{\bmif}{\bifd}
\safemath{\bmig}{\bigd}
\safemath{\bmih}{\bihd}
\safemath{\bmii}{\biid}
\safemath{\bmij}{\bijd}
\safemath{\bmik}{\bikd}
\safemath{\bmil}{\bild}
\safemath{\bmim}{\bimd}
\safemath{\bmin}{\bind}
\safemath{\bmio}{\biod}
\safemath{\bmip}{\bipd}
\safemath{\bmiq}{\biqd}
\safemath{\bmir}{\bird}
\safemath{\bmis}{\bisd}
\safemath{\bmit}{\bitd}
\safemath{\bmiu}{\biud}
\safemath{\bmiv}{\bivd}
\safemath{\bmiw}{\biwd}
\safemath{\bmix}{\bixd}
\safemath{\bmiy}{\biyd}
\safemath{\bmiz}{\bizd}

\safemath{\bmxi}{\bixid}
\safemath{\bmlambda}{\bilambdad}
\safemath{\bmmu}{\bimud}
\safemath{\bmtheta}{\bithetad}
\safemath{\bmphi}{\biphid}
\safemath{\bmdelta}{\bideltad}

\safemath{\bA}{\mathbf{A}}
\safemath{\bB}{\mathbf{B}}
\safemath{\bC}{\mathbf{C}}
\safemath{\bD}{\mathbf{D}}
\safemath{\bE}{\mathbf{E}}
\safemath{\bF}{\mathbf{F}}
\safemath{\bG}{\mathbf{G}}
\safemath{\bH}{\mathbf{H}}
\safemath{\bI}{\mathbf{I}}
\safemath{\bJ}{\mathbf{J}}
\safemath{\bK}{\mathbf{K}}
\safemath{\bL}{\mathbf{L}}
\safemath{\bM}{\mathbf{M}}
\safemath{\bN}{\mathbf{N}}
\safemath{\bO}{\mathbf{O}}
\safemath{\bP}{\mathbf{P}}
\safemath{\bQ}{\mathbf{Q}}
\safemath{\bR}{\mathbf{R}}
\safemath{\bS}{\mathbf{S}}
\safemath{\bT}{\mathbf{T}}
\safemath{\bU}{\mathbf{U}}
\safemath{\bV}{\mathbf{V}}
\safemath{\bW}{\mathbf{W}}
\safemath{\bX}{\mathbf{X}}
\safemath{\bY}{\mathbf{Y}}
\safemath{\bZ}{\mathbf{Z}}

\safemath{\bZero}{\mathbf{0}}
\safemath{\bOne}{\mathbf{1}}
\safemath{\bDelta}{\mathbf{\Delta}}
\safemath{\bLambda}{\mathbf{\UpLambda}}
\safemath{\bPhi}{\mathbf{\Upphi}}
\safemath{\bSigma}{\mathbf{\Upsigma}}
\safemath{\bOmega}{\mathbf{\Upomega}}
\safemath{\bTheta}{\mathbf{\Uptheta}}

\bmdefine{\biAd}{A}
\bmdefine{\biBd}{B}
\bmdefine{\biCd}{C}
\bmdefine{\biDd}{D}
\bmdefine{\biEd}{E}
\bmdefine{\biFd}{F}
\bmdefine{\biGd}{G}
\bmdefine{\biHd}{H}
\bmdefine{\biId}{I}
\bmdefine{\biJd}{J}
\bmdefine{\biKd}{K}
\bmdefine{\biLd}{L}
\bmdefine{\biMd}{M}
\bmdefine{\biOd}{N}
\bmdefine{\biPd}{O}
\bmdefine{\biQd}{P}
\bmdefine{\biRd}{R}
\bmdefine{\biSd}{S}
\bmdefine{\biTd}{T}
\bmdefine{\biUd}{U}
\bmdefine{\biVd}{V}
\bmdefine{\biWd}{W}
\bmdefine{\biXd}{X}
\bmdefine{\biYd}{Y}
\bmdefine{\biZd}{Z}

\bmdefine{\biDelta}{\Delta}
\bmdefine{\biLambda}{\Lambda}
\bmdefine{\biPhi}{\Phi}
\bmdefine{\biSigma}{\Sigma}
\bmdefine{\biOmega}{\Omega}
\bmdefine{\biTheta}{\Theta}

\safemath{\bimA}{\biAd}
\safemath{\bimB}{\biBd}
\safemath{\bimC}{\biCd}
\safemath{\bimD}{\biDd}
\safemath{\bimE}{\biEd}
\safemath{\bimF}{\biFd}
\safemath{\bimG}{\biGd}
\safemath{\bimH}{\biHd}
\safemath{\bimI}{\biId}
\safemath{\bimJ}{\biJd}
\safemath{\bimK}{\biKd}
\safemath{\bimL}{\biLd}
\safemath{\bimM}{\biMd}
\safemath{\bimN}{\biNd}
\safemath{\bimO}{\biOd}
\safemath{\bimP}{\biPd}
\safemath{\bimQ}{\biQd}
\safemath{\bimR}{\biRd}
\safemath{\bimS}{\biSd}
\safemath{\bimT}{\biTd}
\safemath{\bimU}{\biUd}
\safemath{\bimV}{\biVd}
\safemath{\bimW}{\biWd}
\safemath{\bimX}{\biXd}
\safemath{\bimY}{\biYd}
\safemath{\bimZ}{\biZd}

\safemath{\bimDelta}{\biDelta}
\safemath{\bimLambda}{\biLambda}
\safemath{\bimPhi}{\biPhi}
\safemath{\bimSigma}{\biSigma}
\safemath{\bimOmega}{\biOmega}
\safemath{\bimTheta}{\biTheta}

\safemath{\setA}{\mathcal{A}}
\safemath{\setB}{\mathcal{B}}
\safemath{\setC}{\mathcal{C}}
\safemath{\setD}{\mathcal{D}}
\safemath{\setE}{\mathcal{E}}
\safemath{\setF}{\mathcal{F}}
\safemath{\setG}{\mathcal{G}}
\safemath{\setH}{\mathcal{H}}
\safemath{\setI}{\mathcal{I}}
\safemath{\setJ}{\mathcal{J}}
\safemath{\setK}{\mathcal{K}}
\safemath{\setL}{\mathcal{L}}
\safemath{\setM}{\mathcal{M}}
\safemath{\setN}{\mathcal{N}}
\safemath{\setO}{\mathcal{O}}
\safemath{\setP}{\mathcal{P}}
\safemath{\setQ}{\mathcal{Q}}
\safemath{\setR}{\mathcal{R}}
\safemath{\setS}{\mathcal{S}}
\safemath{\setT}{\mathcal{T}}
\safemath{\setU}{\mathcal{U}}
\safemath{\setV}{\mathcal{V}}
\safemath{\setW}{\mathcal{W}}
\safemath{\setX}{\mathcal{X}}
\safemath{\setY}{\mathcal{Y}}
\safemath{\setZ}{\mathcal{Z}}
\safemath{\emptySet}{\varnothing}

\safemath{\colA}{\mathscr{A}}
\safemath{\colB}{\mathscr{B}}
\safemath{\colC}{\mathscr{C}}
\safemath{\colD}{\mathscr{D}}
\safemath{\colE}{\mathscr{E}}
\safemath{\colF}{\mathscr{F}}
\safemath{\colG}{\mathscr{G}}
\safemath{\colH}{\mathscr{H}}
\safemath{\colI}{\mathscr{I}}
\safemath{\colJ}{\mathscr{J}}
\safemath{\colK}{\mathscr{K}}
\safemath{\colL}{\mathscr{L}}
\safemath{\colM}{\mathscr{M}}
\safemath{\colN}{\mathscr{N}}
\safemath{\colO}{\mathscr{O}}
\safemath{\colP}{\mathscr{P}}
\safemath{\colQ}{\mathscr{Q}}
\safemath{\colR}{\mathscr{R}}
\safemath{\colS}{\mathscr{S}}
\safemath{\colT}{\mathscr{T}}
\safemath{\colU}{\mathscr{U}}
\safemath{\colV}{\mathscr{V}}
\safemath{\colW}{\mathscr{W}}
\safemath{\colX}{\mathscr{X}}
\safemath{\colY}{\mathscr{Y}}
\safemath{\colZ}{\mathscr{Z}}

\safemath{\opA}{\mathbb{A}}
\safemath{\opB}{\mathbb{B}}
\safemath{\opC}{\mathbb{C}}
\safemath{\opD}{\mathbb{D}}
\safemath{\opE}{\mathbb{E}}
\safemath{\opF}{\mathbb{F}}
\safemath{\opG}{\mathbb{G}}
\safemath{\opH}{\mathbb{H}}
\safemath{\opI}{\mathbb{I}}
\safemath{\opJ}{\mathbb{J}}
\safemath{\opK}{\mathbb{K}}
\safemath{\opL}{\mathbb{L}}
\safemath{\opM}{\mathbb{M}}
\safemath{\opN}{\mathbb{N}}
\safemath{\opO}{\mathbb{O}}
\safemath{\opP}{\mathbb{P}}
\safemath{\opQ}{\mathbb{Q}}
\safemath{\opR}{\mathbb{R}}
\safemath{\opS}{\mathbb{S}}
\safemath{\opT}{\mathbb{T}}
\safemath{\opU}{\mathbb{U}}
\safemath{\opV}{\mathbb{V}}
\safemath{\opW}{\mathbb{W}}
\safemath{\opX}{\mathbb{X}}
\safemath{\opY}{\mathbb{Y}}
\safemath{\opZ}{\mathbb{Z}}
\safemath{\opZero}{\mathbb{O}}
\safemath{\identityop}{\opI}

\safemath{\veca}{\bma}
\safemath{\vecb}{\bmb}
\safemath{\vecc}{\bmc}
\safemath{\vecd}{\bmd}
\safemath{\vece}{\bme}
\safemath{\vecf}{\bmf}
\safemath{\vecg}{\bmg}
\safemath{\vech}{\bmh}
\safemath{\veci}{\bmi}
\safemath{\vecj}{\bmj}
\safemath{\veck}{\bmk}
\safemath{\vecl}{\bml}
\safemath{\vecm}{\bmm}
\safemath{\vecn}{\bmn}
\safemath{\veco}{\bmo}
\safemath{\vecp}{\bmp}
\safemath{\vecq}{\bmq}
\safemath{\vecr}{\bmr}
\safemath{\vecs}{\bms}
\safemath{\vect}{\bmt}
\safemath{\vecu}{\bmu}
\safemath{\vecv}{\bmv}
\safemath{\vecw}{\bmw}
\safemath{\vecx}{\bmx}
\safemath{\vecy}{\bmy}
\safemath{\vecz}{\bmz}

\safemath{\veczero}{\bmzero}
\safemath{\vecone}{\bmone}
\safemath{\vecxi}{\bmxi}
\safemath{\veclambda}{\bmlambda}
\safemath{\vecmu}{\bmmu}
\safemath{\vectheta}{\bmtheta}
\safemath{\vecphi}{\bmphi}
\safemath{\vecdelta}{\bmdelta}

\safemath{\matA}{\bA}
\safemath{\matB}{\bB}
\safemath{\matC}{\bC}
\safemath{\matD}{\bD}
\safemath{\matE}{\bE}
\safemath{\matF}{\bF}
\safemath{\matG}{\bG}
\safemath{\matH}{\bH}
\safemath{\matI}{\bI}
\safemath{\matJ}{\bJ}
\safemath{\matK}{\bK}
\safemath{\matL}{\bL}
\safemath{\matM}{\bM}
\safemath{\matN}{\bN}
\safemath{\matO}{\bO}
\safemath{\matP}{\bP}
\safemath{\matQ}{\bQ}
\safemath{\matR}{\bR}
\safemath{\matS}{\bS}
\safemath{\matT}{\bT}
\safemath{\matU}{\bU}
\safemath{\matV}{\bV}
\safemath{\matW}{\bW}
\safemath{\matX}{\bX}
\safemath{\matY}{\bY}
\safemath{\matZ}{\bZ}
\safemath{\matzero}{\bmzero}

\safemath{\matDelta}{\bDelta}
\safemath{\matLambda}{\bLambda}
\safemath{\matPhi}{\bPhi}
\safemath{\matSigma}{\bSigma}
\safemath{\matOmega}{\bOmega}
\safemath{\matTheta}{\bTheta}

\safemath{\matidentity}{\matI}
\safemath{\matone}{\matO}

\safemath{\rnda}{A}
\safemath{\rndb}{B}
\safemath{\rndc}{C}
\safemath{\rndd}{D}
\safemath{\rnde}{E}
\safemath{\rndf}{F}
\safemath{\rndg}{G}
\safemath{\rndh}{H}
\safemath{\rndi}{I}
\safemath{\rndj}{J}
\safemath{\rndk}{K}
\safemath{\rndl}{L}
\safemath{\rndm}{M}
\safemath{\rndn}{N}
\safemath{\rndo}{O}
\safemath{\rndp}{P}
\safemath{\rndq}{Q}
\safemath{\rndr}{R}
\safemath{\rnds}{S}
\safemath{\rndt}{T}
\safemath{\rndu}{U}
\safemath{\rndv}{V}
\safemath{\rndw}{W}
\safemath{\rndx}{X}
\safemath{\rndy}{Y}
\safemath{\rndz}{Z}

\safemath{\rveca}{\bimA}
\safemath{\rvecb}{\bimB}
\safemath{\rvecc}{\bimC}
\safemath{\rvecd}{\bimD}
\safemath{\rvece}{\bimE}
\safemath{\rvecf}{\bimF}
\safemath{\rvecg}{\bimG}
\safemath{\rvech}{\bimH}
\safemath{\rveci}{\bimI}
\safemath{\rvecj}{\bimJ}
\safemath{\rveck}{\bimK}
\safemath{\rvecl}{\bimL}
\safemath{\rvecm}{\bimM}
\safemath{\rvecn}{\bimN}
\safemath{\rveco}{\bomO}
\safemath{\rvecp}{\bimP}
\safemath{\rvecq}{\bimQ}
\safemath{\rvecr}{\bimR}
\safemath{\rvecs}{\bimS}
\safemath{\rvect}{\bimT}
\safemath{\rvecu}{\bimU}
\safemath{\rvecv}{\bimV}
\safemath{\rvecw}{\bimW}
\safemath{\rvecx}{\bimX}
\safemath{\rvecy}{\bimY}
\safemath{\rvecz}{\bimZ}

\safemath{\rvecxi}{\bmxi}
\safemath{\rveclambda}{\bmlambda}
\safemath{\rvecmu}{\bmmu}
\safemath{\rvectheta}{\bmtheta}
\safemath{\rvecphi}{\bmphi}

\safemath{\rmatA}{\bimA}
\safemath{\rmatB}{\bimB}
\safemath{\rmatC}{\bimC}
\safemath{\rmatD}{\bimD}
\safemath{\rmatE}{\bimE}
\safemath{\rmatF}{\bimF}
\safemath{\rmatG}{\bimG}
\safemath{\rmatH}{\bimH}
\safemath{\rmatI}{\bimI}
\safemath{\rmatJ}{\bimJ}
\safemath{\rmatK}{\bimK}
\safemath{\rmatL}{\bimL}
\safemath{\rmatM}{\bimM}
\safemath{\rmatN}{\bimN}
\safemath{\rmatO}{\bimO}
\safemath{\rmatP}{\bimP}
\safemath{\rmatQ}{\bimQ}
\safemath{\rmatR}{\bimR}
\safemath{\rmatS}{\bimS}
\safemath{\rmatT}{\bimT}
\safemath{\rmatU}{\bimU}
\safemath{\rmatV}{\bimV}
\safemath{\rmatW}{\bimW}
\safemath{\rmatX}{\bimX}
\safemath{\rmatY}{\bimY}
\safemath{\rmatZ}{\bimZ}

\safemath{\rmatDelta}{\bimDelta}
\safemath{\rmatLambda}{\bimLambda}
\safemath{\rmatPhi}{\bimPhi}
\safemath{\rmatSigma}{\bimSigma}
\safemath{\rmatOmega}{\bimOmega}
\safemath{\rmatTheta}{\bimTheta}

%% file: macros/standard-macros.tex
\usepackage{amssymb}
\usepackage{amsfonts}
\usepackage{mathrsfs}
\usepackage{xspace}
\usepackage{bm}
\usepackage{fancyref}
\usepackage{textcomp}

\usepackage{multirow}
\usepackage{stmaryrd}

\newenvironment{textbmatrix}{	\setlength{\arraycolsep}{2.5pt}%
								\big[\begin{matrix}}{\end{matrix}\big]%
								\raisebox{0.08ex}{\vphantom{M}}}

\def\be{\begin{equation}}
\def\ee{\end{equation}}
\def\een{\nonumber \end{equation}}
\def\mat{\begin{bmatrix}}
\def\emat{\end{bmatrix}}
\def\btm{\begin{textbmatrix}}
\def\etm{\end{textbmatrix}}

\def\ba#1\ea{\begin{align}#1\end{align}}
\def\bas#1\eas{\begin{align*}#1\end{align*}}
\def\bs#1\es{\begin{split}#1\end{split}}
\def\bg#1\eg{\begin{gather}#1\end{gather}}
\def\bml#1\eml{\begin{multline}#1\end{multline}}
\def\bi#1\ei{\begin{itemize}#1\end{itemize}}

\newcommand{\lefto}{\mathopen{}\left}

\DeclareMathOperator{\sign}{sign}			
\DeclareMathOperator{\Prob}{\opP}			

\newcommand{\abs}[1]{\lefto\lvert#1\right\rvert}		

\safemath{\dirac}{\delta}					
\safemath{\krond}{\dirac}					

\safemath{\upto}{\uparrow}
\safemath{\downto}{\downarrow}
\safemath{\iu}{j}							
\safemath{\ev}{\lambda}						
\safemath{\hilseqspace}{l^{2}}				
\newcommand{\banachfunspace}[1]{\setL^{#1}}	
\safemath{\hilfunspace}{\banachfunspace{2}}	

\safemath{\SNR}{\textit{SNR}} 				
\safemath{\PAR}{\textit{PAR}} 				
\safemath{\No}{N_0}							
\safemath{\Es}{E_s}							
\safemath{\Eb}{E_b}							
\safemath{\EbNo}{\frac{\Eb}{\No}}
\safemath{\EsNo}{\frac{\Es}{\No}}

\DeclareMathOperator{\CHop}{\ensuremath{\opH}} 
\safemath{\tvir}{\rndh_{\CHop}}				
\safemath{\tvtf}{\rndl_{\CHop}}				
\safemath{\spf}{\rnds_{\CHop}}				
\safemath{\bff}{H_{\CHop}}					

\safemath{\ircf}{r_{h}}						
\safemath{\tftvcf}{r_{s}}					
\safemath{\tfcf}{r_{l}}						
\safemath{\bfcf}{r_{H}}						

\safemath{\tcorr}{c_h}						
\safemath{\scf}{c_{s}}						
\safemath{\tfcorr}{c_{l}}					
\safemath{\fcorr}{c_{H}}						

\safemath{\mi}{I}							
\safemath{\capacity}{C}						

\safemath{\normal}{\mathcal{N}}			
\safemath{\jpg}{\mathcal{CN}}			
\safemath{\mchain}{\leftrightarrow}		

\safemath{\dB}{\,\mathrm{dB}}
\safemath{\dBm}{\,\mathrm{dBm}}
\safemath{\Hz}{\,\mathrm{Hz}}
\safemath{\kHz}{\,\mathrm{kHz}}
\safemath{\MHz}{\,\mathrm{MHz}}
\safemath{\GHz}{\,\mathrm{GHz}}
\safemath{\s}{\,\mathrm{s}}
\safemath{\ms}{\,\mathrm{ms}}
\safemath{\mus}{\,\mathrm{\text{\textmu}s}}
\safemath{\ns}{\,\mathrm{ns}}
\safemath{\ps}{\,\mathrm{ps}}
\safemath{\meter}{\,\mathrm{m}}
\safemath{\mm}{\,\mathrm{mm}}
\safemath{\cm}{\,\mathrm{cm}}
\safemath{\m}{\,\mathrm{m}}
\safemath{\W}{\,\mathrm{W}}
\safemath{\mW}{\, \mathrm{mW}}
\safemath{\J}{\,\mathrm{J}}
\safemath{\K}{\,\mathrm{K}}
\safemath{\bit}{\,\mathrm{bit}}
\safemath{\nat}{\,\mathrm{nat}}

\safemath{\define}{\triangleq}			

\safemath{\equivalent}{\sim}
\safemath{\distas}{\sim}					
\safemath{\sdiff}{\Delta}				

\safemath{\reals}{\mathbb{R}}
\safemath{\positivereals}{\reals_{+}}
\safemath{\integers}{\mathbb{Z}}
\safemath{\posint}{\integers_{+}}
\safemath{\naturals}{\mathbb{N}}
\safemath{\posnaturals}{\naturals_{+}}
\safemath{\complexset}{\mathbb{C}}
\safemath{\rationals}{\mathbb{Q}}

\newcommand*{\fancyrefapplabelprefix}{app}		
\newcommand*{\fancyrefthmlabelprefix}{thm}		
\newcommand*{\fancyreflemlabelprefix}{lem}		
\newcommand*{\fancyrefcorlabelprefix}{cor}		
\newcommand*{\fancyrefdeflabelprefix}{def}		
\newcommand*{\fancyrefproplabelprefix}{prop}		
\newcommand*{\fancyrefexmpllabelprefix}{exmpl}
\newcommand*{\fancyrefalglabelprefix}{alg}		
\newcommand*{\fancyreftbllabelprefix}{tbl}		

\frefformat{vario}{\fancyrefseclabelprefix}{Sec.~#1}
\frefformat{vario}{\fancyrefthmlabelprefix}{Thm.~#1}
\frefformat{vario}{\fancyreftbllabelprefix}{Tbl.~#1}
\frefformat{vario}{\fancyreflemlabelprefix}{Lem.~#1}
\frefformat{vario}{\fancyrefcorlabelprefix}{Cor.~#1}
\frefformat{vario}{\fancyrefdeflabelprefix}{Def.~#1}
\frefformat{vario}{\fancyreffiglabelprefix}{Fig.~#1}
\frefformat{vario}{\fancyrefapplabelprefix}{App.~#1}
\frefformat{vario}{\fancyrefeqlabelprefix}{(#1)}
\frefformat{vario}{\fancyrefproplabelprefix}{Prop.~#1}
\frefformat{vario}{\fancyrefexmpllabelprefix}{Ex.~#1}
\frefformat{vario}{\fancyrefalglabelprefix}{Alg.~#1}

%% file: macros/defs.tex
\safemath{\dictab}{[\,\dicta\,\,\dictb\,]}

\safemath{\ysig}{\bmy}
\safemath{\ysighat}{\hat{\ysig}}
\safemath{\ysigdim}{M}
\safemath{\xsig}{\bmx}
\safemath{\xsigdim}{N}
\safemath{\nx}{n_x}
\safemath{\zsig}{\bmz}
\safemath{\zsigdim}{\ysigdim}
\safemath{\rsig}{\bmr}
\safemath{\Adict}{\bA}
\safemath{\Adicttilde}{\widetilde{\Adict}}
\safemath{\Adictdim}{\outputdim\times\xsigdim}
\safemath{\avec}{\bma}
\safemath{\avectilde}{\tilde{\avec}}
\safemath{\Bdict}{\bB}
\safemath{\Bdicttilde}{\widetilde{\Bdict}}
\safemath{\Cdict}{\bC}
\safemath{\cvec}{\bmc}
\safemath{\Ddict}{\bD}
\safemath{\Ddictdim}{\ysigdim\times\xsigdim}
\safemath{\dvec}{\bmd}
\safemath{\Ddicttilde}{\widetilde{\bD}}
\safemath{\Bonb}{\bB}
\safemath{\bvec}{\bmb}
\safemath{\Bonbdim}{\ysigdim\times\ysigdim}
\safemath{\noise}{\bmn}
\safemath{\noisedim}{\ysigim}
\safemath{\err}{\bme}
\safemath{\errdim}{\ysigdim}
\safemath{\errset}{\setE}
\safemath{\nerr}{n_e}
\safemath{\delop}{\bP_\errset}
\safemath{\delopc}{\bP_{{\errset}^c}}

\safemath{\cplxi}{\imath}
\safemath{\cplxj}{\jmath}

\safemath{\dict}{\matD}
\safemath{\inputdim}{N}		
\safemath{\outputdim}{M}		
\safemath{\sparsity}{S}	
\safemath{\inputdimA}{{N_a}}	
\safemath{\inputdimB}{{N_b}}	
\safemath{\elemA}{{n_a}}	
\safemath{\elemB}{{n_b}}	
\safemath{\resA}{\matR_a}	
\safemath{\resB}{\matR_b}	
\safemath{\subD}{\matS} 
\safemath{\subA}{\matS_a} 
\safemath{\subB}{\matS_b} 
\safemath{\dicta}{\matA} 	
\safemath{\dictb}{\matB} 	
\safemath{\hollowS}{H}
\safemath{\hollowA}{H_a}
\safemath{\hollowB}{H_b}
\safemath{\cross}{Z}
\safemath{\coh}{\mu_d}			
\safemath{\coha}{\mu_a}			
\safemath{\cohb}{\mu_b}			
\safemath{\mubs}{\nu}	
\safemath{\cohm}{\mu_m} 
\safemath{\dictset}{\setD}	
\safemath{\dictsetp}{\dictset(\coh,\coha,\cohb)}	
\safemath{\dictsetgen}{\dictset_\text{gen}}
\safemath{\dictsetgenp}{\dictsetgen(\coh)}
\safemath{\dictsetonb}{\dictset_\text{onb}}
\safemath{\dictsetonbp}{\dictsetonb(\coh)}

\safemath{\leftside}{U}
\safemath{\rightsideA}{R_a}
\safemath{\rightsideB}{R_b}

\safemath{\indexS}{\setI_S} 

\safemath{\na}{n_a}			
\safemath{\nb}{n_b}			
\safemath{\coeffa}{p_i}	
\safemath{\coeffb}{q_j}	
\safemath{\seta}{\setP}		
\safemath{\setb}{\setQ}     
\safemath{\setw}{\setW}	
\safemath{\setz}{\setZ}	
\safemath{\cola}{\veca}		
\safemath{\colb}{\vecb}		
\safemath{\cold}{\vecd}		
\safemath{\inputvec}{\vecx} 	
\safemath{\error}{\vece}	
\safemath{\noiseout}{\vecz} 	
\safemath{\inputvecel}{x}
\safemath{\inputveca}{\vecx_a}
\safemath{\inputvecb}{\vecx_b}
\safemath{\outputvec}{\vecy}	
\safemath{\lambdamin}{\lambda_{\mathrm{min}}}

\safemath{\elltwo}{\ell_2}
\safemath{\ellone}{\ell_1}
\safemath{\ellzero}{\ell_0}
\safemath{\ellinf}{\ell_\infty}
\safemath{\ellinftilde}{\ell_{\widetilde\infty}}
\safemath{\licard}{Z(\coh,\coha,\cohb)}
\safemath{\xsol}{\hat{x}}
\safemath{\xbord}{x_b}		
\safemath{\xstat}{x_s}		
\safemath{\xstatLone}{\tilde{x}_s}
\safemath{\order}{\mathcal{O}} 
\safemath{\scales}{\Theta} 
\safemath{\ones}{\mathbf{1}} 
\safemath{\zeroes}{\mathbf{0}} 
\safemath{\thlone}{\kappa(\coh,\cohb)} 
\safemath{\constoneA}{\delta} 
\safemath{\constoneB}{\epsilon} 
\safemath{\nlarge}{L}				   
\safemath{\sumlarge}{S_\nlarge}
\safemath{\maxlarger}{P_\nlarge}	   
\safemath{\Pzero}{\textrm{P0}}	
\safemath{\Pone}{\textrm{P1}}
\safemath{\vecfir}{\vecw}			 
\safemath{\vecsec}{\vecz}
\safemath{\elvecfir}{w}              
\safemath{\elvecsec}{z}				 
\safemath{\nlargefir}{n}
\safemath{\normout}{\gamma}
\safemath{\auxfun}{h}
\safemath{\supp}{\textrm{supp}}

\safemath{\indexa}{\ell}
\safemath{\indexb}{r}
\safemath{\indexc}{i}
\safemath{\indexd}{j}

\safemath{\project}{P}

%% file: chapters/intro.tex

\section{Introduction}

Fifth-generation (5G) wireless systems include ultra-reliable low-latency communication (URLLC)~\cite{Lin2022} to support applications in industry automation, autonomous driving, and telemedicine.
Achieving low latency requires codes with short blocklengths for which polar codes with successive cancellation list (SCL) decoding are suitable candidates~\cite{IT11}. However, SCL decoding is highly sequential, which results in decoder implementations that achieve long latency and rather poor area efficiency~\cite{PG17,CT21}.
In contrast, low-density parity-check (LDPC) codes~\cite{RG62} are adopted in virtually all modern communication systems because message-passing (MP)-based decoders achieve high throughput and excellent area- and energy-efficiency while delivering near capacity performance in the long blocklength regime~\cite{MK99}.
However, MP-based LDPC decoders typically do \emph{not} perform well with short-blocklength codes, and despite their potential use for URLLC applications, little is known about short-blocklength LDPC codes that achieve low error rates with MP-based decoding and corresponding high-throughput, low-latency hardware implementations.
\subsection{Contributions}
\begin{figure}[tp]
\centering
\includegraphics[width=0.95\columnwidth]{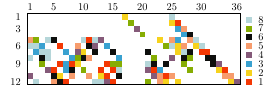}
\caption{Base graph of the proposed short-blocklength binary quasi-cyclic LDPC code with lifting factor $Z=8$. Each square represents a cyclically shifted $8\times8$ identity matrix with the shift amount indicated by the corresponding color. The entire matrix is used for blocklengths of $288$ coded bits; for blocklengths $224$ and $160$ coded bits, we remove the first $8$ and $16$ block columns, respectively.}
\label{fig:code_matrix}
\end{figure}
We propose a novel short-blocklength binary quasi-cyclic (QC) LDPC code~\cite{MF04} that supports three rates.
Our code is optimized to achieve good error-correcting performance with MP-based decoding, which enables efficient hardware implementations that achieve low latency. 
To demonstrate our code's effectiveness, we propose an ASIC implementation of a flooding-schedule MP-based decoder that processes each decoding iteration in a fully parallel fashion.
To improve error-rate performance, we employ machine-learning-optimized edge-adaptive normalized min-sum (ANMS) check-node updates~\cite{EN18}; to improve energy efficiency, we utilize early termination (ET); and to improve throughput without sacrificing latency, we use pipeline interleaving~\cite[Sec.~3.7.6]{HK} to process two independent codewords simultaneously.
Our decoder has been fabricated in GlobalFoundries' 22FDX technology, occupies an area of only 0.44\,mm$^2$, and measurements reveal information throughputs of $9$, $16$ and $22$\,Gb/s at record latencies of $13.78$, $16.06$, and $17.52$\,ns for code rates 1$/$2, 2$/$3, and 3$/$4, respectively.
\subsection{Relevant Prior Work}

The literature describes numerous LDPC decoder designs, which mostly target long blocklengths.
Reference~\cite{RG18} provides place-and-route results for a fully parallel LDPC decoder supporting a blocklength of $2048$ bits, which unrolls all iterations to achieve extremely high throughput. 
Reference~\cite{ZZ10} targets the same blocklength and achieves state-of-the-art area and energy efficiency at the expense of reduced throughput with a partially parallel decoder architecture. 
However, only little is known about decoders targeting short LDPC codes, with the exception of~\cite{MM18,AV24}. 
Similar to~\cite{MM18,AV24}, our decoder is rate-configurable, and similar to~\cite{MM18}, we use early termination (ET) to save power and pipeline interleaving to improve throughput.  
In contrast to~\cite{MM18,AV24},  we propose a novel LDPC code of even shorter blocklength and a decoder ASIC that achieves the lowest latency reported in the literature. In addition, our decoder achieves excellent area efficiency and error rate performance and is, thus, suitable for URLLC.

Polar codes and corresponding SCL decoders are known to perform well with short blocklengths. In contrast to the state-of-the-art polar decoder implementations from~\cite{PG17,CT21}, we implement a hardware-friendly LDPC decoder, achieving orders of magnitude lower latency. Furthermore, our decoder produces soft outputs on the coded bits, which is useful for hybrid automatic repeat request or iterative detection and decoding.

Specifically for URLLC applications, there exists the recently proposed block orthogonal sparse superposition (BOSS) decoder~\cite{DK24}; this decoder achieves low throughput and is area inefficient. 
In comparison to all of the discussed state-of-the-art designs, our decoder achieves the shortest decoding latency while reaching the highest throughput among all designs targeting short blocklengths; see \fref{sec:results} for a comparison.

%% file: chapters/vlsi.tex

\section{VLSI Architecture}
\begin{figure}[tp]
\centering
\includegraphics[width=0.95\columnwidth]{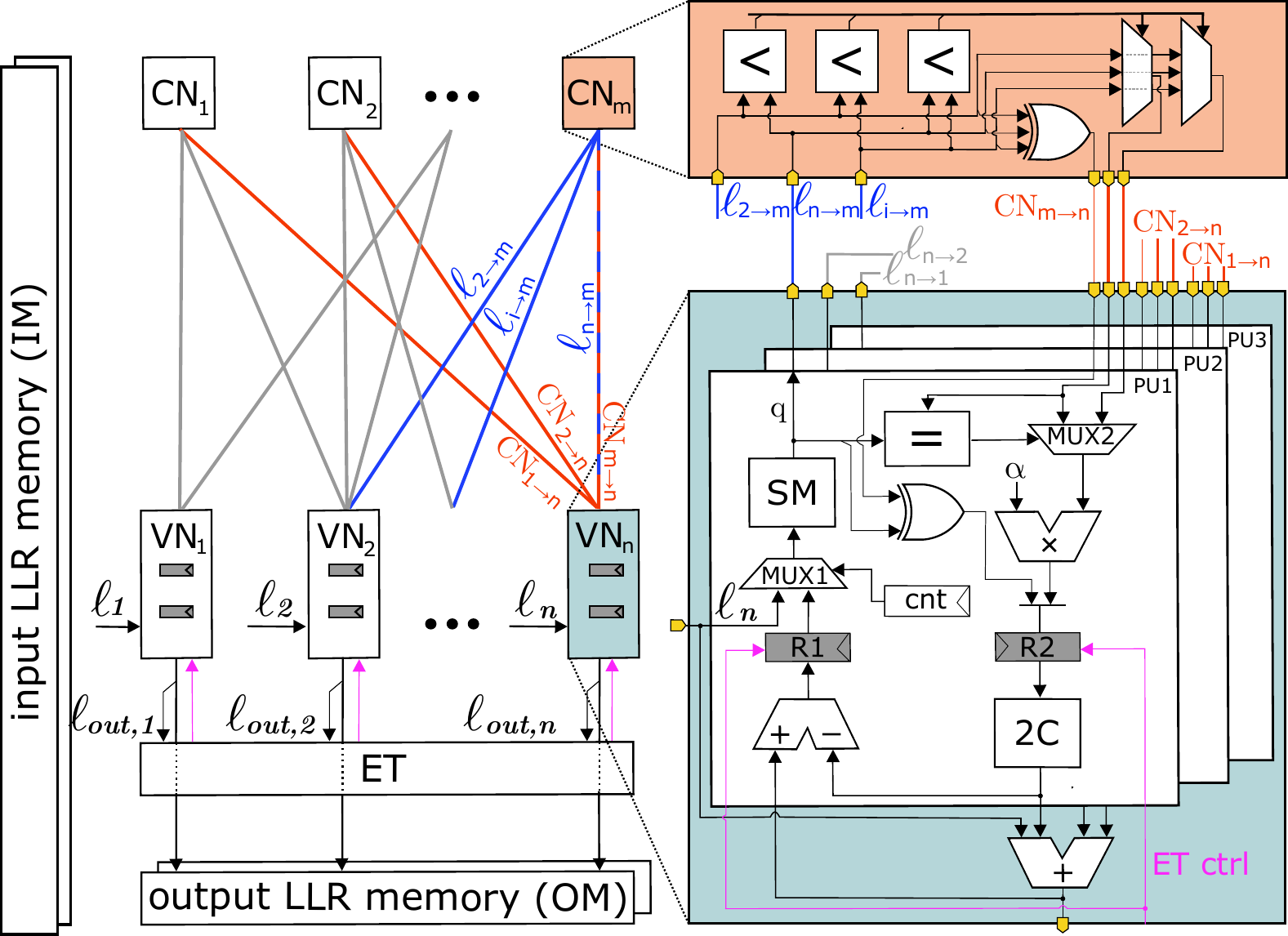}
\caption{Architecture of the proposed SB-LDPC decoder. The input/output LLR memories contains two sets of LLR values associated to two independent codewords. In this example, VN$_\textnormal{n}$ contains three PUs as it connects to three CNs (CN$_1$, CN$_2$, and CN$_\textnormal{m}$); each PU contains two pipeline registers, which enable us to process two codewords at the same time. The ET block triggers pipeline register freezing in case a valid codeword is found.}
\vspace{-0.1cm}
\label{fig:architecture}
\end{figure}

\subsection{Architecture Overview}

To achieve high throughput at low latency, we implement the flooding schedule in a fully parallel fashion per MP iteration. 
The left side of \fref{fig:architecture} provides an architecture overview. Our decoder consists of $N=288$ VN blocks (one for each code bit of the longest blocklength) and $M=96$ CN blocks.
VN processing and CN processing together carry out  NMS~\cite{Chen02} MP as in \fref{eq:check_node_eq}.
To improve throughput, we deploy pipeline interleaving~\cite[Sec.~3.7.6]{HK} with pipeline registers in the VN-blocks; this allows simultaneous processing of two independent codewords while shortening the critical path. 
To improve energy efficiency, we utilize ET, which dynamically freezes the appropriate pipeline stages to suppress switching activity.

\subsection{Architecture Details}
The right side of \fref{fig:architecture} provides the architecture details.
Each VN block contains a multi-operand adder and processing units (PUs). Each PU is responsible for one of the VN-CN connections of the VN block (determined by the PC matrix in \fref{fig:code_matrix}).
Starting at pipeline register R1, we first convert the VN-to-CN messages (or the LLR values $\ell_i$ in the first iteration) from two's complement to sign magnitude values $q$ (SM block).
Then, these messages are sent to the appropriate CN blocks, which extract the smallest and second smallest absolute values of the messages coming from the connected VN blocks (see blue connections in \fref{fig:architecture}). Each CN block also computes the associated parity check by XOR-ing all incoming sign bits. The CN processing is facilitated by the sign magnitude format.
Back in the PU, the value $q$ is compared to the minimum from the incoming CN. If they match, then MUX2 passes the second-smallest absolute value; if not, MUX2 passes the minimum. The selected value is then scaled using the predefined weight as in \fref{eq:check_node_eq}; since all of the weights are determined offline, only multiplications with constants are necessary.
The resulting sign-magnitude value is then stored in pipeline register R2. 
After R2, the message is converted to two's complement (2C block) in preparation for summation. 
Each VN block then sums all the CN-to-VN messages previously processed by the internal PUs, including the associated input LLR $\ell_i$, obtaining the intrinsic LLR $\ell_{\textnormal{out},i}$ associated with code bit $c_i$.
The intrinsic LLRs are fed back to the PUs, where they are converted into extrinsic messages and stored in R1. 
Decoding is repeated for $I_\textnormal{max}$ iterations. The ET block uses the intrinsic LLR’s sign bits to compute the PC.
If ET is triggered for one of the two interleaved codewords (pink signal in \fref{fig:architecture}), then the pipeline registers associated with the unterminated codeword keep their outputs for one additional clock cycle to avoid switching activity associated with the early-terminated codeword. If ET is triggered for both codewords, then all pipeline registers are frozen.
To minimize the BLER implementation loss (cf.~\fref{fig:bler}), we use fixed-point numbers with $4$ integer bits, $2$ fractional bits, and $1$ sign bit. The multi-operand adder inside the VNs uses one additional integer bit.